\documentclass[runningheads]{llncs}
\usepackage{graphicx}
\usepackage{xr-hyper}
\usepackage{hyperref}
\usepackage{multirow}
\makeatletter
\newcommand*{\addFileDependency}[1]{
  \typeout{(#1)}
  \@addtofilelist{#1}
  \IfFileExists{#1}{}{\typeout{No file #1.}}
}
\makeatother

\newcommand*{\myexternaldocument}[1]{
    \externaldocument{#1}
    \addFileDependency{#1.tex}
    \addFileDependency{#1.aux}
}
\myexternaldocument{supMaterial}

\begin{document}
\title{DeepEdit: Deep Editable Learning for Interactive Segmentation of 3D Medical Images}

\titlerunning{DeepEdit: Deep Editable Learning}




\author{Andres Diaz-Pinto\inst{1,6} \and
Pritesh Mehta\inst{1} \and
Sachidanand Alle\inst{6} \and
Muhammad Asad\inst{1} \and
Richard Brown\inst{1} \and
Vishwesh Nath\inst{6} \and
Alvin Ihsani\inst{6} \and
Michela Antonelli\inst{1} \and
Daniel Palkovics\inst{2} \and
Csaba Pinter\inst{3} \and
Ron Alkalay\inst{4} \and
Steve Pieper\inst{5} \and
Holger R. Roth\inst{6} \and
Daguang Xu\inst{6} \and
Prerna Dogra\inst{6} \and
Tom Vercauteren\inst{1} \and
Andrew Feng\inst{6} \and
Abood Quraini\inst{6} \and 
Sebastien Ourselin\inst{1} \and
M. Jorge Cardoso\inst{1}}
\authorrunning{Diaz-Pinto et al.}
%
\institute{School of Biomedical Engineering \& Imaging Sciences, King’s College London. London, UK. \email{\{andres.diaz-pinto,m.jorge.cardoso\}@kcl.ac.uk} \and
Department of Periodontology, Semmelweis University, Hungary \and
EBATINCA, S.L. Canary Islands, Spain. \and
Beth Israel Deaconess Medical Center, MA, USA. \and
Isomics, Inc., MA, USA. \and
NVIDIA Santa Clara, CA, USA. 
}


%
\maketitle 
\begin{abstract}
Automatic segmentation of medical images is a key step for diagnostic and interventional tasks. However, achieving this requires large amounts of annotated volumes, which can be tedious and time-consuming task for expert annotators. In this paper, we introduce DeepEdit, a deep learning-based method for volumetric medical image annotation, that allows automatic and semi-automatic segmentation, and click-based refinement. DeepEdit combines the power of two methods: a non-interactive (i.e. automatic segmentation using nnU-Net, UNET or UNETR) and an interactive segmentation method (i.e. DeepGrow), into a single deep learning model. It allows easy integration of uncertainty-based ranking strategies (i.e. aleatoric and epistemic uncertainty computation) and active learning. We propose and implement a method for training DeepEdit by using standard training combined with user interaction simulation. Once trained, DeepEdit allows clinicians to quickly segment their datasets by using the algorithm in auto segmentation mode or by providing clicks via a user interface (i.e. 3D Slicer, OHIF). We show the value of DeepEdit through evaluation on the PROSTATEx dataset for prostate/prostatic lesions and the Multi-Atlas Labeling Beyond the Cranial Vault (BTCV) dataset for abdominal CT segmentation, using state-of-the-art network architectures as baseline for comparison. DeepEdit could reduce the time and effort annotating 3D medical images compared to DeepGrow alone. Source code is available at \url{https://github.com/Project-MONAI/MONAILabel}





\end{abstract}
\section{Introduction}
\label{sec:introduction}

Inspired by the landmark contributions of 2D U-Net\cite{Ronneberger2015}, 3D U-Net\cite{Cicek2016}, and V-Net\cite{Milletari2016}, Convolutional Neural Networks (CNN) have become high-performing methods for automatic segmentation of medical images \cite{Isensee2020,He2021_journal,Hatamizadeh2021_journal}. Medical image segmentation challenges, such as the Medical Segmentation Decathlon (\textit{MSD}) \cite{Antonelli2021_journal}, has helped steer methodological innovations and performance improvements for CNN-based methods. At the time of writing, one of the first positions on the live leaderboard~\footnote{\url{https://decathlon-10.grand-challenge.org/evaluation/challenge/leaderboard/}} for \textit{MSD} is held by the nnU-Net \cite{Isensee2020}, a segmentation pipeline based on U-Net that automatically configures to any new medical image segmentation task. More recently, transformer-based \cite{Vaswani2017} networks introduced by Hatamizadeh et. al. (Swin UNETR \cite{hatamizadeh2022swin_journal} and UNETR \cite{Hatamizadeh2021_journal}), have further improved on nnUNET's performance, achieving state-of-the-art performance on the \textit{MSD} segmentation tasks. 


Despite their outstanding performance, automatic segmentation algorithms have not yet reached the desired level of performance needed for certain clinical applications \cite{Sakinis2019}. In particular, automatic segmentation accuracy can be impacted by patient variation, acquisition differences, image artifacts \cite{Zhao2013} and limited amount of training data. In an attempt to address these challenges, interactive segmentation methods that accept user guidance to improve segmentation have been proposed \cite{Shi2000,Grady2005,Boykov2006,Akkus2015}. Normalized cuts \cite{Shi2000}, random walks \cite{Grady2005}, graph-cuts \cite{Boykov2006}, and geodesics \cite{Akkus2015} have been proposed for interactive segmentation using bounding-box or scribbles-based user interactions. However, a major limitation of these classical methods is that they only succeed in addressing simpler segmentation problems where objects have clear structural boundaries, and require extensive user interaction for more complex segmentation cases containing ambiguity in object boundaries \cite{Sakinis2019}.


A number of deep learning-based interactive segmentation methods based have been proposed for improving the robustness of interactive image segmentation \cite{Xu2016,Agustsson2019}. In \cite{Xu2016}, user foreground and background clicks were converted into euclidean distance maps, and subsequently learned from as additional input channels to a CNN. Inspired by the aforementioned studies and other incremental works, interactive methods for medical image segmentation based on deep learning have been recently proposed \cite{Wang2018,Sakinis2019,Wang2019}. In \cite{Wang2018}, a bounding-box and scribble-based CNN segmentation pipeline was proposed, whereby a user-provided bounding box is first used to assist the CNN in foreground segmentation. This was followed by image-specific fine-tuning using user-provided scribbles. Due to the inclusion of user interaction within a CNN, this method provided greater robustness and accuracy than state-of-the-art for segmenting previously unseen objects, while also using fewer user interactions than existing interactive segmentation methods. In contrast, Sakinis et al.~\cite{Sakinis2019} proposed a click-based method, motivated in part by the work of \cite{Xu2016}. In their work, Gaussian-smoothed foreground and background clicks were added as input channels to an encoder-decoder CNN. Experiments on multiple-organ segmentation in CT volumes showed that their method delivers 2D segmentations in a fast and reliable manner, generalizes well to unseen structures, and accurately segments organs with few clicks. An alternate method that first performs an automatic CNN segmentation step, followed by an optional refinement through user clicks or scribbles, was proposed by \cite{Wang2019}. Their method, named DeepIGeoS, achieved substantially improved performance compared to automatic CNN on 2D placenta and 3D brain tumour segmentation, and higher accuracy with fewer interactions than traditional interactive segmentation methods.

Automatic and semi-automatic segmentation methods are available as part of open-source software packages for medical imaging analysis: ITK-SNAP \cite{py06nimg} which offers semi-automatic active contour segmentation \cite{Kass1988}; 3D Slicer \cite{Fedorov2012} and MITK \cite{Nolden2013} offer automatic, boundary-points-based \cite{Maninis2018}; DeepGrow~\cite{Sakinis2019} segmentation through the NVIDIA Clara AI-Assisted Annotation Extension; as well as other classic semi-automatic segmentation methods such as region growing \cite{Adams1994} and level sets \cite{Osher1988}.

We propose DeepEdit, a method that combines an automatic and a semi-automatic approach for 3D medical images into a single deep learning-based model. DeepEdit has three working modes: first, it can be used in click-free inference mode (similar to a regular segmentation network), providing fully-automatic segmentation predictions which can be used as a form of initialisation; second, it allows users to provide clicks to initialise and guide a semi-automatic segmentation model; lastly, given an initial segmentation, DeepEdit can be used to refine and improve the initial prediction by providing editing clicks. DeepEdit training process is similar to the algorithm proposed by Sakinis et al.~\cite{Sakinis2019} (DeepGrow) - Gaussian-smoothed clicks for all labels and background are generated and added as input to the backbone CNN, but removes the minimum-click limitation of DeepGrow. Contrary to DeepGrow, our proposed DeepEdit model allows the prediction of an automatic segmentation-based initialisation without user-provided clicks, which can then be further edited by providing clicks. Lastly, the proposed model can also be used for multi-label segmentation problems, allowing the user to generate/segment all labels simultaneously instead of one label at a time.

The flexibility offered by embedding these three functionalities (auto segmentation, semi-automatic segmentation and label refinement) allows DeepEdit to be integrated into an active learning pipeline. For instance, it could be used in automatic segmentation mode for aleatoric and/or epistemic uncertainty computation to rank unlabeled images (See Fig.~\ref{general_schema_deepedit}(b)).

In order to show the performance of DeepEdit, we present applications for single and multiple label segmentation for annotating the datasets: prostate, prostatic lesion, and abdominal organ segmentation. 





\section{Proposed Method}

\begin{figure}[h!]
\centering
\includegraphics[width=0.99\textwidth]{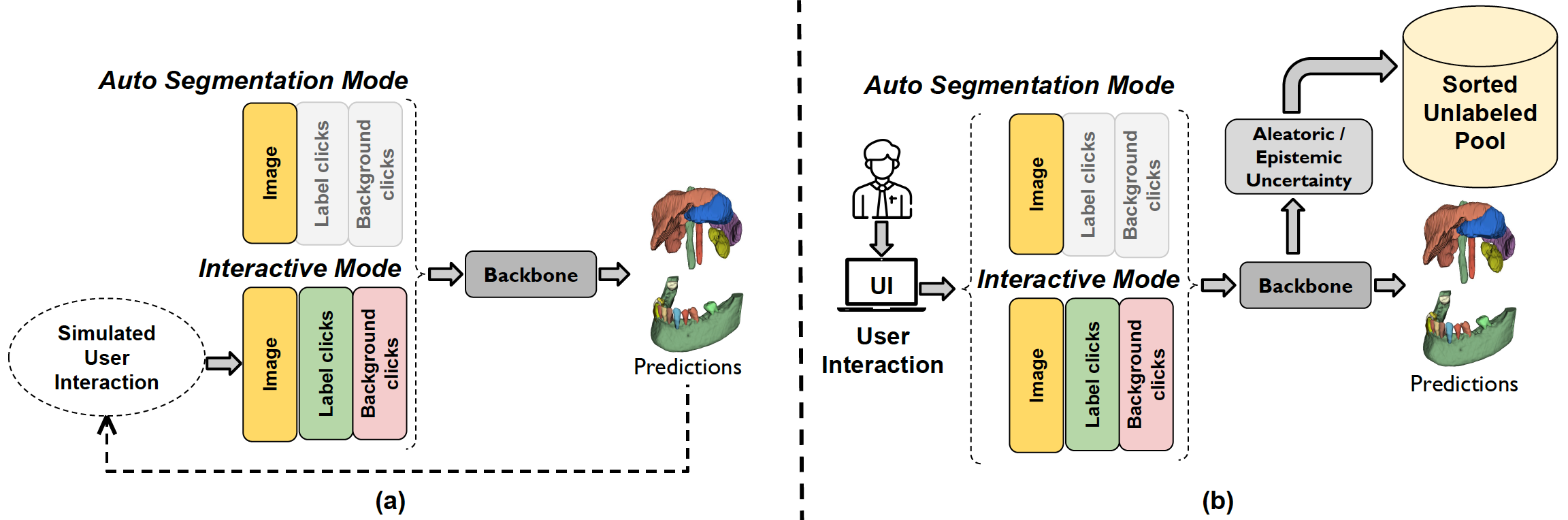} 
\caption{\textbf{General schema of the DeepEdit:} (a) Training and (b) Inference Mode. DeepEdit training process consists of two modes: the automatic segmentation mode and the interactive mode. Simulated clicks for all labels plus background are added to a backbone network as input channels . Input tensor could be either the image with zero-tensors (automatic segmentation mode) or the image with tensors representing label clicks and background clicks provided by the user (interactive mode).}
\label{general_schema_deepedit}
\end{figure}

The DeepEdit architecture is based on a backbone that can be any segmentation network (i.e. UNET, nnU-Net \cite{Isensee2020}, UNETR, SwinUNETR\cite{hatamizadeh2022swin_journal}). The main difference resides in how this backbone is trained and the number of channels in the input tensor. For training, the input tensor could be either the image with zeroed tensors (automatic segmentation mode) or the image with tensors representing label and background clicks provided by the user (interactive mode). In Fig.~\ref{general_schema_deepedit}, DeepEdit is presented in its training and inference mode. 

As shown in Fig.~\ref{general_schema_deepedit}, DeepEdit can integrate an active learning strategy in which the trained model is used to rank the unlabelled volumes from the most uncertain to the least uncertain. Every time the expert annotator fetches an image, DeepEdit present the one with more uncertainty, allowing the model to learn from the most challenging cases first.

\subsection{User interaction and simulated clicks}

Our proposed method embeds three approaches: automatic segmentation, semi-automatic segmentation and interactive segmentation. This means, for some iterations, DeepEdit is trained click-free, and for others is trained as DeepGrow (clicks are simulated and included in the input tensor as extra channels). As DeepGrow relies on clicks provided by a user or agent, we simulated those following the similar approach presented in Sakinis' work - voxels where clicks are located are set to one and smoothed with a Gaussian filter. This is done for the positive label, if single label task or, in general, for all labels and background.



\subsection{Training DeepEdit}

As previously mentioned, the training process of the DeepEdit algorithm involves click-free iterations and iterations with simulated clicks. As shown in Fig.~\ref{general_schema_deepedit}(a), the input of the network is a concatenation of multiple tensors: the image, a tensor containing clicks simulated for each label and a tensor containing clicks simulated for the background. Our proposed algorithm mixes two types of training loops: a) click-free training iterations - meaning that for some iterations, the tensors representing the labels and background clicks are zeros (training for the automatic segmentation); b) simulated-click based training iterations - where labels and background clicks are simulated and placed in the tensors. We sample from a uniform distribution with probability $p$ to determine which iteration is click-free and which one uses simulated clicks. An additional hyper-parameter, the number of simulated clicks per iteration, is set by the user as a function of the task complexity. These two training loops allow DeepEdit to be used fully automatically, semi-automatic, and as a segmentation refinement approach. We developed all these new transforms for click simulation and mixed training in MONAI \cite{Sakinis2019,MONAI2020}.


\section{Experimental Results}

In order to demonstrate the flexibility and value of DeepEdit, and the impact of the number of simulated clicks, a set of experiments were performed on the PROSTATEx and Multi Atlas Labeling Beyond The Cranial Vault (BTCV) datasets. 
We present the impact of the number of clicks in the prostate, prostatic lesion, and abdominal organs (BTCV dataset) segmentation. For both single and multilabel segmentation experiments, we used a learning rate of 1e-4, batch size equal to 1 and Adam optimizer. The following MONAI transforms were used to train and validate DeepEdit: intensity normalization, random flipping (vertically and horizontally), random shift intensity and random rotation.


All our experiments have been implemented using the MONAI Core library \cite{MONAI2020} (version 0.8.1) and MONAI Label platform (version 0.3.1). All source code for DeepEdit algorithm and Active Learning strategies have been made publicly available and documented at \url{https://github.com/Project-MONAI/MONAILabel} as part of the MONAI Label repository.

\subsection{Prostate Segmentation Tasks}

DeepEdit applications were built for whole prostate segmentation and prostatic lesion segmentation. Experiments were run using the PROSTATEx Challenge training dataset, \cite{LitjensPX2017}, hereby referred to as the PROSTATEx dataset. For both single- and multi-label tasks, experiments were conducted to compare the segmentation performance of DeepEdit as the hyperparameter controlling the number of training iterations with zero simulated clicks is varied. We compared DeepEdit on different click-free training iterations: DeepEdit-0 (equivalent to DeepGrow), DeepEdit-0.25 and DeepEdit-0.50, meaning that 0, 25 and 50 percent of the training iterations were click-free. 
Ten-fold cross-validation was performed for both tasks. Segmentation quality was assessed using the Dice coefficient. As in \cite{Sakinis2019}, segmentation performance at inference time was assessed using simulated clicks instead of user mouse-clicks to objectively assess how segmentation quality improves as clicks are added; segmentation performance was assessed at 0, 1, 5, and 10 simulated inference clicks. The presented results are an average of three repetitions to  account for variability in simulated inference click placement.

\subsubsection{Whole Prostate Segmentation}
The whole prostate segmentation task concerns the segmentation of the prostate on T2-weighted MRI (T2WI). Eleven patients from the PROSTATEx dataset were excluded due to inconsistencies between T2WI and the ground-truth segmentations, leaving a total of 193 patients for use in experiments. 

T2WI were pre-processed by resampling to a common resolution of 0.5mm $\times$ 0.5mm $\times$ 3.0mm, normalization using per-image whitening, and cropping/padding to a common size of 320$\times$320 $\times$32. 

\begin{figure}[b!]
\centering
\includegraphics[width=0.8\columnwidth]{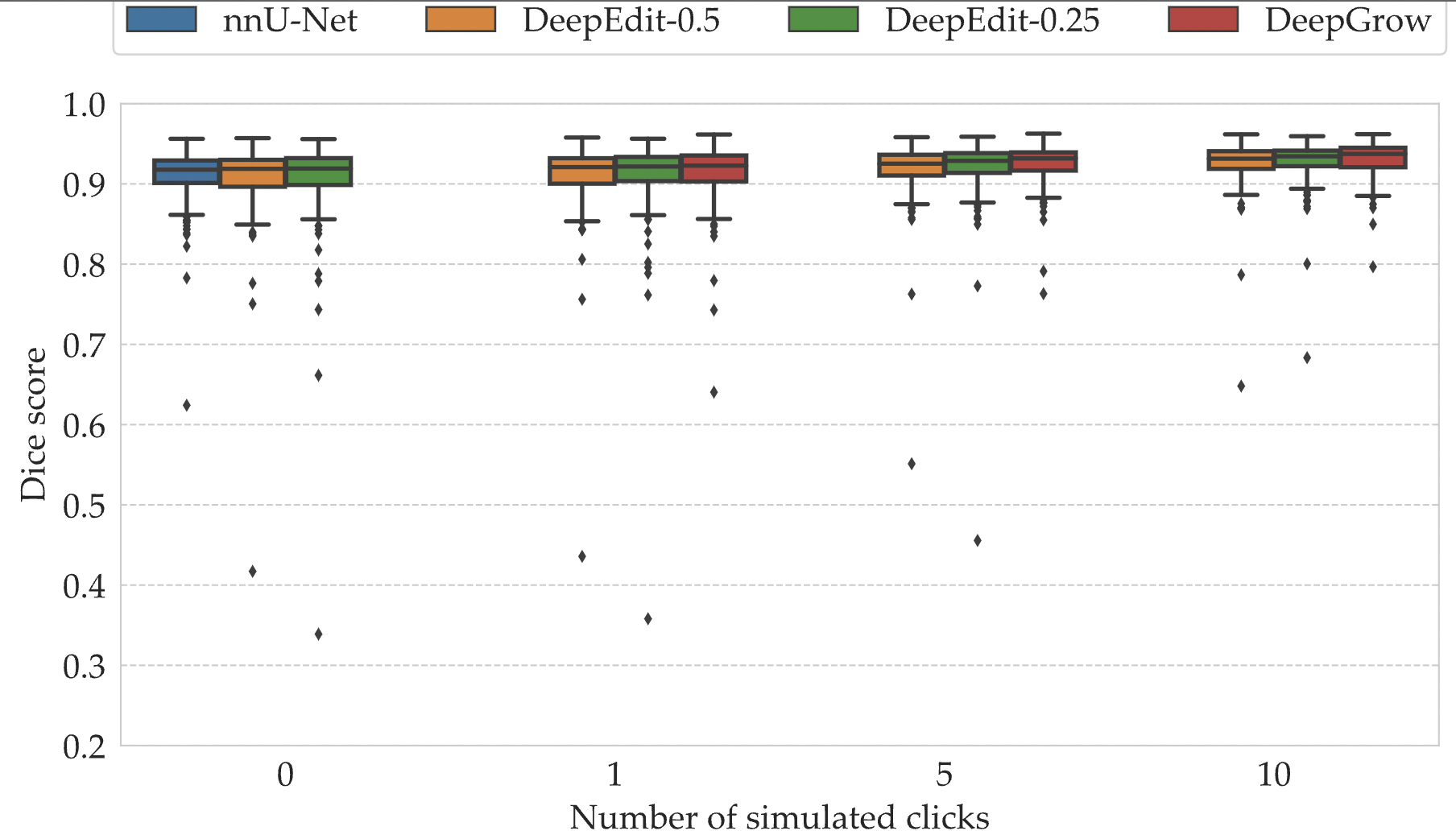}
\caption{\textbf{Whole prostate segmentation:} Dice score box plots for the 193 PROSTATEx dataset patients used in the ten-fold cross-validation, for DeepEdit-0 (equivalent to DeepGrow), DeepEdit-0.25, and DeepEdit-0.5.}
\label{fig:whole_prostate_dice}
\vspace{-20pt}
\end{figure}

A comparison of DeepEdit-0 (equivalent to DeepGrow), DeepEdit-0.25, and DeepEdit-0.5 is shown in Table \ref{tab:whole_prostate_results_table}. Furthermore, the distributions of Dice scores are shown in Fig.~\ref{fig:whole_prostate_dice}. DeepEdit-0.5 was found to have the highest click-free mean Dice score (0.908), while DeepEdit-0 gave the highest mean Dice scores at 1 to 10 simulated inference clicks.

\begin{table}[h!]
\centering
\scalebox{0.80}{
\begin{tabular}{c|c|cccc}
\hline
\multirow{2}{*}{Scheme}                                                                     & \multirow{2}{*}{Model} & \multicolumn{4}{c}{Number of simulated clicks}                                                                                                                           \\ \cline{3-6} 
                                                                                            &                        & \multicolumn{1}{c|}{0}                        & \multicolumn{1}{c|}{1}                        & \multicolumn{1}{c|}{5}                        & 10                       \\ \hline
Fully automatic                                                                             & nnU-Net \cite{Isensee2020}                & \multicolumn{1}{l|}{\textbf{0.910$\pm$0.033}} & \multicolumn{1}{c|}{-}                        & \multicolumn{1}{c|}{-}                        & -                        \\ \hline
\multirow{2}{*}{\begin{tabular}[c]{@{}c@{}}Automatic \\ + interactive editing\end{tabular}} & DeepEdit-0.25          & \multicolumn{1}{c|}{0.908$\pm$0.054}          & \multicolumn{1}{c|}{0.912$\pm$0.049}          & \multicolumn{1}{c|}{0.921$\pm$0.041}          & 0.929$\pm$0.026          \\ \cline{2-6} 
                                                                                            & DeepEdit-0.5           & \multicolumn{1}{c|}{0.908$\pm$0.046}          & \multicolumn{1}{c|}{0.911$\pm$0.044}          & \multicolumn{1}{c|}{0.919$\pm$0.035}          & 0.926$\pm$0.028          \\ \hline
Fully interactive                                                                           & DeepEdit-0 (DeepGrow)             & \multicolumn{1}{c|}{0.907$\pm$0.041}          & \multicolumn{1}{c|}{\textbf{0.915$\pm$0.035}} & \multicolumn{1}{c|}{\textbf{0.926$\pm$0.024}} & \textbf{0.932$\pm$0.020} \\ \hline
\end{tabular}
}
\caption{Whole prostate segmentation mean Dice scores $\pm$ one standard deviation for the 193 PROSTATEx dataset patients used in the ten-fold cross-validation, for nnU-Net, DeepEdit-0 (equivalent to DeepGrow), DeepEdit-0.25, and DeepEdit-0.5. The highest mean Dice in each column is shown in bold.}
\label{tab:whole_prostate_results_table}
\vspace{-20pt}
\end{table}

\subsubsection{Prostatic Lesion Segmentation}
The prostatic lesion segmentation task concerns the segmentation of lesions within the prostate using T2WI, apparent diffusion coefficient (ADC) map, and computed high b-value diffusion-weighted MRI (DWI). Since our experiments were conducted using the PROSTATEx dataset, we used the PROSTATEx definition of a lesion, i.e., a prostatic lesion is defined as any area of suspicion attributed to a Prostate Imaging-Reporting and Data System (PI-RADS) score by the expert radiologist (anonymous clinician) who read and reported PROSTATEx dataset cases; all lesions in the PROSTATEx dataset were scored PI-RADS $\geq 2$. Four patients from the PROSTATEx dataset were excluded due to not containing contoured lesions in the ground truth (the assessment metrics would have been undefined), leaving a total of 200 patients with a total of 299 lesions for use in experiments.


A b-value, $b=2000$, was selected for computing high b-value DWI; computed b2000 (Cb2000) DWI were generated using DWI acquired at lower b-values, extrapolated by assuming a monoexponential model for the per-voxel observed signal. ADC map and Cb2000 DWI were registered to T2WI to account for voluntary/involuntary patient movement between acquisitions and differences in resolution. T2WI and Cb2000 DWI were normalised by dividing voxel intensities by the interquartile mean of central gland (CG) voxel intensities \cite{Mehta2021a}; ADC maps were not normalised as they contain a quantitative measurement. T2WI, ADC map, and Cb2000 DWI were resampled to a common resolution of 0.5 mm $\times$ 0.5 mm $\times$ 3 mm. Then, whole prostate masks were used to crop the prostate region on all MR modalities; a margin was applied in each direction to reduce the likelihood of prostate tissue being discarded. Next, a cropping/padding transformation was used to ensure a common spatial size of 256 $\times$ 256 $\times$ 32.

\begin{figure}[h!]
\centering
\includegraphics[width=0.75\columnwidth]{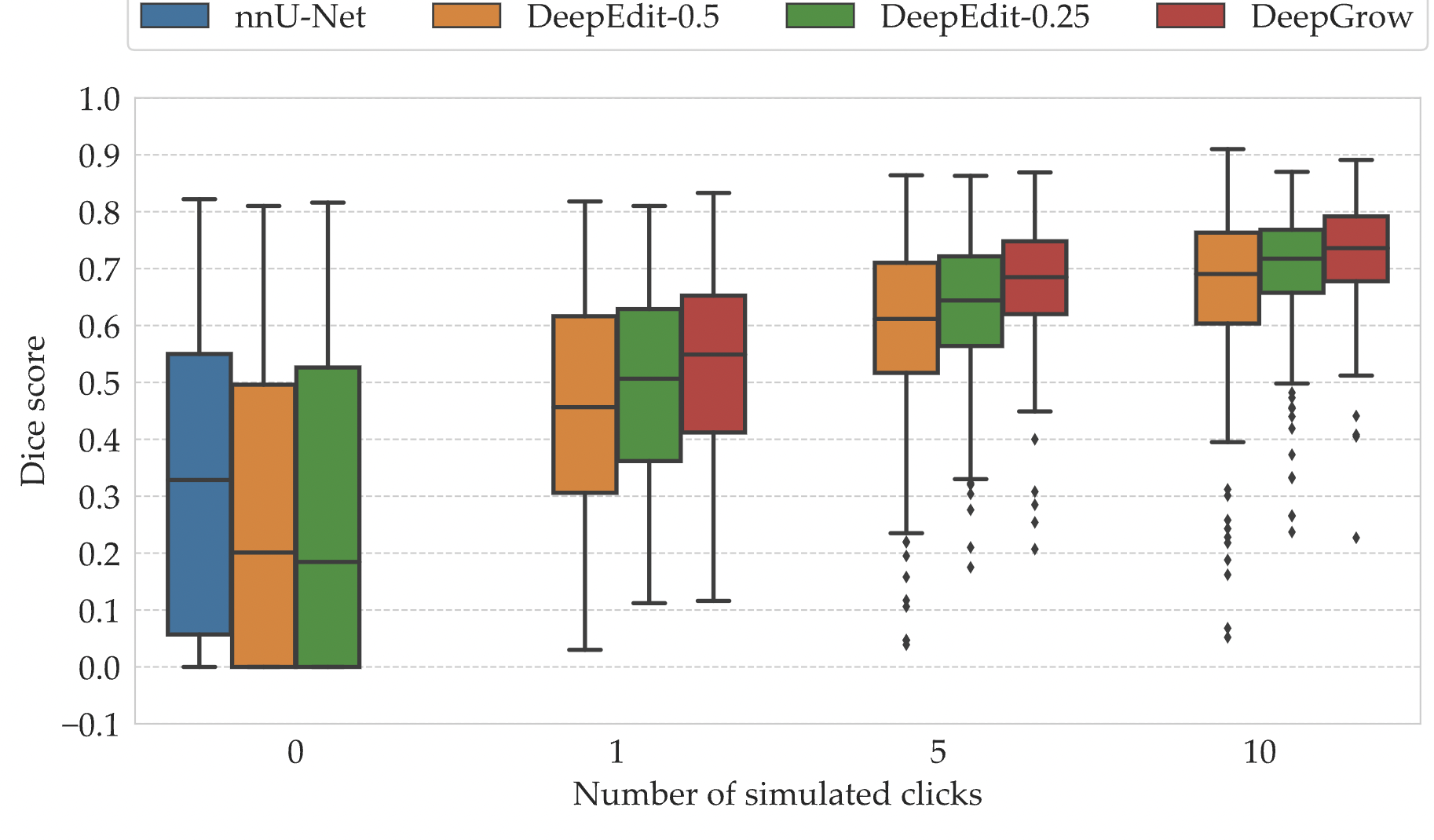}
\caption{\textbf{Prostatic lesion segmentation:} Dice score box plots for the 200 PROSTATEx dataset patients used in the ten-fold cross-validation, for DeepEdit-0 (equivalent to DeepGrow), DeepEdit-0.25, and DeepEdit-0.5.}
\label{fig:prostatic_lesion_dice}
\end{figure}

A comparison of DeepEdit-0 (equivalent to DeepGrow), DeepEdit-0.25, and DeepEdit-0.5 is shown in Table \ref{tab:prostatic_lesion_results_table}. Furthermore, the distributions of Dice scores are shown in Fig \ref{fig:prostatic_lesion_dice}. As in the whole prostate segmentation task, DeepEdit-0.5 gave the highest click-free mean Dice score (0.272), while DeepEdit-0 (equivalent to DeepGrow) gave the highest mean Dice scores at 1 to 10 simulated inference clicks.

\begin{table}[t!]
\centering
\scalebox{0.80}{
\begin{tabular}{c|c|cccc}
\hline
\multirow{2}{*}{Scheme}                                                                     & \multirow{2}{*}{Model} & \multicolumn{4}{c}{Number of simulated clicks}                                                                                                                           \\ \cline{3-6} 
                                                                                            &                        & \multicolumn{1}{c|}{0}                        & \multicolumn{1}{c|}{1}                        & \multicolumn{1}{c|}{5}                        & 10                       \\ \hline
Fully automatic                                                                             & nnU-Net \cite{Isensee2020}                & \multicolumn{1}{l|}{\textbf{0.332$\pm$0.254}} & \multicolumn{1}{c|}{-}                        & \multicolumn{1}{c|}{-}                        & -                        \\ \hline
\multirow{2}{*}{\begin{tabular}[c]{@{}c@{}}Automatic \\ + interactive editing\end{tabular}} & DeepEdit-0.25          & \multicolumn{1}{c|}{0.268$\pm$0.271}          & \multicolumn{1}{c|}{0.498$\pm$0.174}          & \multicolumn{1}{c|}{0.632$\pm$0.130}          & 0.697$\pm$0.114          \\ \cline{2-6} 
                                                                                            & DeepEdit-0.5           & \multicolumn{1}{c|}{0.272$\pm$0.266}          & \multicolumn{1}{c|}{0.453$\pm$0.197}          & \multicolumn{1}{c|}{0.592$\pm$0.163}          & 0.663$\pm$0.145          \\ \hline
Fully interactive                                                                           & DeepEdit-0 (DeepGrow)  & \multicolumn{1}{c|}{0.166$\pm$0.254}          & \multicolumn{1}{c|}{\textbf{0.527$\pm$0.166}} & \multicolumn{1}{c|}{\textbf{0.670$\pm$0.111}} & \textbf{0.723$\pm$0.095} \\ \hline
\end{tabular}
}
\caption{Prostatic lesion segmentation mean Dice scores $\pm$ one standard deviation for the 200 PROSTATEx dataset patients used in the ten-fold cross-validation, for nnU-Net, DeepEdit-0 (equivalent to DeepGrow), DeepEdit-0.25, and DeepEdit-0.5. The highest mean Dice in each column is shown in bold.}
\label{tab:prostatic_lesion_results_table}
\vspace{-20pt}
\end{table}

\subsection{Abdominal Organ Segmentation}

A second set of experiments using the UNETR \cite{Hatamizadeh2021_journal} as backbone were performed on the BTCV dataset. For this, we used 23 images for training, 6 for validation, and an image size of $128\times128\times128$. As the previous analysis, we compared DeepEdit trained with 0\% click-free training iterations (equivalent to DeepGrow), 25\% click-free training iterations, and 50\% click-free training iterations. 



\begin{table}[h!]
\centering
\scalebox{0.8}{
\begin{tabular}{c|c|cccc|cccc}
\hline
\multirow{3}{*}{Scheme}                                                                     & \multirow{3}{*}{Model} & \multicolumn{4}{c|}{Single label}                                                                                                & \multicolumn{4}{c}{Multilabel}                                                                                                   \\ \cline{3-10} 
                                                                                            &                        & \multicolumn{4}{c|}{Number of simulated clicks}                                                                                  & \multicolumn{4}{c}{Number of simulated clicks}                                                                                   \\ \cline{3-10} 
                                                                                            &                        & \multicolumn{1}{c|}{0}              & \multicolumn{1}{c|}{1}              & \multicolumn{1}{c|}{5}              & 10             & \multicolumn{1}{c|}{0}              & \multicolumn{1}{c|}{1}              & \multicolumn{1}{c|}{5}              & 10             \\ \hline
Fully automatic                                                                             & UNETR                  & \multicolumn{1}{c|}{\textbf{0.919}} & \multicolumn{1}{c|}{-}              & \multicolumn{1}{c|}{-}              & -              & \multicolumn{1}{c|}{\textbf{0.911}} & \multicolumn{1}{c|}{-}              & \multicolumn{1}{c|}{-}              & -              \\ \hline
\multirow{2}{*}{\begin{tabular}[c]{@{}c@{}}Automatic \\ + interactive editing\end{tabular}} & DeepEdit-0.25          & \multicolumn{1}{c|}{0.874}          & \multicolumn{1}{c|}{\textbf{0.902}} & \multicolumn{1}{c|}{0.895}          & 0.895          & \multicolumn{1}{c|}{0.901}          & \multicolumn{1}{c|}{0.887}          & \multicolumn{1}{c|}{0.900}          & 0.906          \\ \cline{2-10} 
                                                                                            & DeepEdit-0.5           & \multicolumn{1}{c|}{0.835}          & \multicolumn{1}{c|}{0.850}          & \multicolumn{1}{c|}{0.864}          & 0.876          & \multicolumn{1}{c|}{0.875}          & \multicolumn{1}{c|}{\textbf{0.895}} & \multicolumn{1}{c|}{0.899}          & 0.905          \\ \hline
Fully interactive                                                                           & DeepEdit-0 (DeepGrow)  & \multicolumn{1}{c|}{0.897}          & \multicolumn{1}{c|}{0.899}          & \multicolumn{1}{c|}{\textbf{0.913}} & \textbf{0.931} & \multicolumn{1}{c|}{0.892}          & \multicolumn{1}{c|}{0.892}          & \multicolumn{1}{c|}{\textbf{0.914}} & \textbf{0.926} \\ \hline
\end{tabular}
}
\caption{\textbf{Dice scores for single and multilabel segmentation on the validation set using the BTCV dataset.} For single label we used the spleen organ and multilabel we used spleen, liver, and left and right kidneys. We show the results obtained for 0, 1, 5, and 10 clicks simulated during validation. Highest Dice scores in each column are shown in bold.}
\label{tab:single_multilabel_results}
\vspace{-20pt}
\end{table}

In Table \ref{tab:single_multilabel_results}, we show the obtained results on the validation set for 0, 1, 5, and 10 simulated clicks. As a fair comparison, we trained and validated a UNETR and the DeepEdit using the same images, same transforms and for the same number of epochs (200). 

As it is shown in Table \ref{tab:single_multilabel_results}, any DeepEdit configuration performs slightly better than the UNETR on the validation set when simulated clicks are provided.

Additional qualitative results obtained from DeepEdit are presented in the supplementary material. We show how DeepEdit could also be applied on two additional clinical problems: segmentation of metastatic spines and teeth segmentation for treatment planning in reconstructive periodontal surgery.




\section{Conclusion}

In this study, we introduce DeepEdit, a method that enables an uncertainty-driven active learning workflow for labelling medical images using a framework that combines deep learning-based automatic segmentation and interactive edits. Compared to previous interactive approaches, DeepEdit can be easily integrated into any 3D medical segmentation pipeline that includes active learning strategies. Using DeepEdit, biologists/clinicians can 1) obtain an automatic segmentation that can later be modified or refined by providing clicks through a user interface (e.g., 3D Slicer, OHIF), or 2) provide clicks to get a segmentation (semi-automatic segmentation). This could significantly reduce the time clinicians/biologists spend on annotating more datasets, which translates in less cost and effort spent on this process. 




%
%
%
%
\bibliographystyle{ieeetr} 
\bibliography{refs}

\end{document}